\documentclass{article}
\usepackage{spconf,amsmath,graphicx}
\usepackage{url}
\usepackage{multirow}
\usepackage{color, colortbl}
\definecolor{LightCyan}{rgb}{0.88,1,1}
\usepackage{xcolor}
\usepackage{caption}
\usepackage{subcaption}
\usepackage{amsmath}
\usepackage{float}
\usepackage{url}
\usepackage{graphicx}
\usepackage{hyperref}
\title{Learning Disentangled Phone and Speaker Representations in a Semi-Supervised VQ-VAE Paradigm}
\name{Jennifer Williams$^1$, Yi Zhao$^2$, Erica Cooper$^2$, Junichi Yamagishi$^2$\sthanks{This work was partially supported by the EPSRC Centre for Doctoral Training in Data Science, funded by the UK Engineering and Physical Sciences Research Council (grant EP/L016427/1) and University of Edinburgh; and by a JST CREST Grant (JPMJCR18A6, VoicePersonae project), Japan. The numerical calculations were carried out on the TSUBAME 3.0 supercomputer at the Tokyo Institute of Technology.}}
\address{$^1$The Centre for Speech Technology Research, University of Edinburgh, UK\\
$^2$National Institute of Informatics, Japan\\
j.williams@ed.ac.uk,$\{$zhaoyi,ecooper,jyamagis$\}$@nii.ac.jp}
\begin{document}
%\ninept
%
\maketitle
\begin{abstract}We present a new approach to disentangle speaker voice and phone content by introducing new components to the VQ-VAE architecture for speech synthesis. The original VQ-VAE does not generalize well to unseen speakers or content. To alleviate this problem, we have incorporated a speaker encoder and speaker VQ codebook that learns global speaker characteristics entirely separate from the existing sub-phone codebooks. We also compare two training methods: self-supervised with global conditions and semi-supervised with speaker labels. Adding a speaker VQ component improves objective measures of speech synthesis quality (estimated MOS, speaker similarity, ASR-based intelligibility) and provides learned representations that are meaningful. Our speaker VQ codebook indices can be used in a simple speaker diarization task and perform slightly better than an x-vector baseline. Additionally, phones can be recognized from sub-phone VQ codebook indices in our semi-supervised VQ-VAE better than self-supervised with global conditions. 

\end{abstract}
\section{Introduction}
%Many open questions remain regarding which types of information from the speech signal need to be explicitly represented, as well as how this is best accomplished. 
Rich representations of speaker identity, such as \textit{x}-vectors~\cite{snyder2018x}, \textit{i}-vectors~\cite{dehak2010front}, and learnable dictionary encodings (LDE)~\cite{cai2018exploring}, are commonplace in many speech technologies ranging from speaker and language recognition, to text-to-speech (TTS) synthesis~\cite{cooper2020zero}, and voice conversion (VC)~\cite{ding2019group}. Representations of speaker identity are also known to contain unwanted information not necessarily related to speaker voice because they often encode extra information such as recording environment, speaker emotion, speaking style, and lexical content. Machine learning methods can remove these unwanted informational factors, so that speaker representations can be made more pure. However, these techniques discard informational factors rather than storing them in separate representations. In addition, evaluation for disentanglement success has been limited to performing ad-hoc classification tasks, whereby a lowered classification accuracy indicates that information has been removed~\cite{williams2019disentangling, raj2019probing,peri2020empirical}. The ability to successfully \textbf{disentangle} and \textbf{retain} information from the speech signal has far-reaching benefits for speech processing. For example, voice conversion and voice privacy benefits from the ability to retain content while changing speaker characteristics. In text-to-speech synthesis, it is desirable to separately control factors of content, speaker, and expressive style. 

In this paper, we present an end-to-end solution to disentangle sub-phone content and global-level speaker characteristics by building upon the original vector-quantized variational autoencoder (VQ-VAE) used in voice conversion. While VQ-VAE learns a type of sub-phone representation~\cite{van2017neural}, the system cannot generalize well to unseen speakers or unseen content. We present our method to learn two different VQ codebooks at the same time while producing synthetic speech that is highly intelligible and with high speaker similarity in unseen conditions. We also demonstrate that the learned representations can be used in downstream tasks: phone recognition from VQ sub-phone codes, and speaker diarization from VQ speaker codes.

\section{Related Work}
%In speech synthesis, a well-known representation is the ``global style token'' (GST) first proposed by~\cite{wang2018style}. GSTs are embeddings learned during TTS training and subsequently applied during inference to control speaking style output. GSTs do not require prosodic labels, and the style representations are separate from lexical content. However, the representations are highly specific to TTS applications: they are not reusable. And some style tokens capture different types of \textit{noise} rather than speaking styles. 

Recent work from~\cite{ebbers2020adversarial} aims to disentangle content from speaker for voice conversion tasks. While their technique does not necessarily rely on speaker labels, it does use vocal tract length perturbation (VTLP) to facilitate regularization during adversarial disentanglement. To facilitate content encoding, the encoder input uses instance normalization for each log-mel-band, for each input signal, separately. As mentioned earlier, the disentanglement success is only defined in terms of ad-hoc classification tasks. The approach does not create representations that are re-usable outside of the system. 

In the VQ-VAE paradigm, learned VQ spaces result in rich continuous-valued embeddings as well as their corresponding discrete code indices. As embeddings, the learned representations can be used to condition a decoder, such as WaveRNN, locally or globally. It was previously shown that a large VQ space can learn to represent phones and sub-phones when trained with speech audio~\cite{van2017neural}. More recently, VQ-VAE has been successfully modified to learn two separate VQ spaces at the same time: F0 and sub-phones. Incorporating an additional encoder for F0 greatly improved speech synthesis quality for Japanese and Mandarin. However, the speaker representation was a one-hot encoded speaker vector or a separately obtained LDE vector~\cite{zhao2020improved}. Our work builds on this effort by learning the speaker VQ space during training and using it for global conditioning in the WaveRNN decoder. In addition, we demonstrate that the VQ code indices are very useful in other downstream tasks.

\section{Architectures}
\begin{figure}
\centering
   \includegraphics[width=8.5cm]{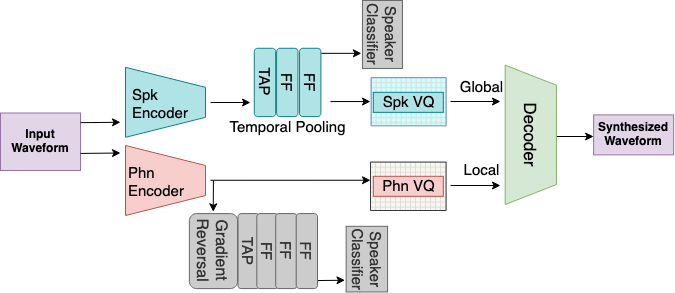}
   \caption{System overview: Speaker VQ encodes global conditions with temporal average pooling layer (TAP). Phone VQ encodes local conditions. Colored components are described in Section 3.2, and gray components in Section 3.3}
   \label{fig:overview}
\end{figure}

\subsection{Original Self-Supervised VQ-VAE}
We used the PyTorch implementation of the original VQ-VAE~\cite{van2017neural} for audio. This system is self-supervised during training as it does not require labeled data, and the learning process is governed by three terms in the loss function which balance the encoder, decoder, and VQ dictionary (Equation~\ref{eq:main_loss}).
\begin{equation}
L = L_{R} + \alpha L_{VQ} + \beta L_{C}
\label{eq:main_loss}
\end{equation}
First, the $L_{R}$ term is the reconstruction loss, defined as $-\log p(x|z_{q}(x))$ which is the negative log likelihood of decoder output $x$ given the output of the encoder $z(x)$ after quantization $q$. The second term $L_{VQ}$ is the VQ objective, defined as $\rVert {\mbox{sg}}[z_e(x)]-e\rVert_2^2$ and it is an $l_2$ loss which guides VQ embedding vectors $e$ towards encoder output $z_e(x)$. The ${\mbox{sg}}$ term is a stop-gradient operator which effectively creates a non-updated constant. The purpose of the VQ term is to ensure that embeddings are also guided by reconstruction loss. Finally, the $L_{C}$ term is a commitment loss defined as $\rVert z_e(x) - {\mbox{sg}}[e] \rVert_2^2$ to ensure that the encoder commits to a VQ embedding vector $e$ and constrains how the VQ space is utilized. In our VQ-VAE implementation, the size of the sub-phone VQ codebook is 512 and each VQ embedding vector is 128-dims. The WaveRNN decoder is implemented as described in~\cite{zhao2020improved}, and it performs local conditioning (using sub-phone VQ embeddings) as well as global conditioning (using speaker one-hot vectors). 

\subsection{Self-Supervised VQ-VAE + Global Conditioning}
Building upon the open-source implementation provided by~\cite{zhao2020improved}, we likewise added an encoder and VQ codebook component to the original VQ-VAE architecture. However, our addition was a VQ component that models global conditions, rather than their F0 VQ component for local conditions. To transform the encoder output into global conditions, we added a temporal average pooling layer and two feed-forward layers between the encoder and VQ dictionary (colored components of Figure~\ref{fig:overview}). Our dual-encoder VQ-VAE is also self-supervised, and we have adjusted the loss function to account for the additional global terms (Equation~\ref{eq:global_cond_loss}).
\begin{equation}
L = L_{R} + \alpha(L_{VQl} + L_{Cl}) + \beta(L_{VQg} + L_{Cg})
\label{eq:global_cond_loss}
\end{equation}
Similar to the original VQ-VAE defined earlier, our dual-encoder model is based on a weighted linear combination of losses. The local features ($L_{VQl}$, $L_{Cl}$) and the global features ($L_{VQg}$, $L_{Cg}$), while retaining the original reconstruction loss. In this system (\textbf{+Global VQ}), the sub-phone VQ codebook remains the same, while we have added a global VQ codebook of size 256, and each VQ embedding is 128-dims.

\subsection{Semi-Supervised VQ-VAE + Speaker Codebook}
Different from the self-supervised training method in Section 3.2, we explicitly provided speaker labels as additional information. Using speaker labels, we created two semi-supervised variants of the dual-encoder model and global conditions from Section 3.2. First, we added an auxiliary speaker classifier to the global encoder components, using VCTK speaker labels as ground truth. We experimented with two different loss functions for the speaker classifier, and we refer them as \textbf{+Speaker label, Softmax} for the softmax loss, and \textbf{+Speaker label, A-softmax} for the angular softmax loss. In each case, the classifier loss is added to the loss function of Equation~\ref{eq:global_cond_loss}.
Finally, we added an additional auxiliary adversarial component to the phone encoder, which consists of a gradient reversal layer~\cite{ganin2016domain}, feed-forward layers, and a speaker classifier. The purpose of this adversarial component was to encourage disentanglement of phone and speaker information. The speaker classifier for the adversarial component used softmax loss. As before, concerning the non-adversarial speaker classification task, we compare the softmax loss (\textbf{+Adversarial loss, Softmax}) and the angular softmax loss (\textbf{+Adversarial loss, A-softmax}). The semi-supervised components are shown in gray in Figure 1.

%\begin{figure*}[ht!]
%\begin{subfigure}[b]{0.31\textwidth}
%  \centering
%   \includegraphics[width=\textwidth]{figures/plt_mos2_estimation.png}
%   \caption{Predicted MOS quality.}
%   \label{fig:mos2}
%   \end{subfigure}
%\begin{subfigure}[b]{0.31\textwidth}
%  \centering
%   \includegraphics[width=\textwidth]{figures/plt_speaker_similarity.png}
%   \caption{Speaker \textit{x}-vector cosine similarity.}
%   \label{fig:speaker_cosine}
%   \end{subfigure}
%   \begin{subfigure}[b]{0.33\textwidth}
%\centering
%   \includegraphics[width=\textwidth]{figures/plt_WER.png}
%   \caption{Watson ASR Word Error Rate (WER).}
%   \label{fig:wer}
%\end{subfigure}
%   \caption{Speech synthesis quality estimation for four testing conditions on %VCTK data (MOS, speaker similarity, WER) }
%   \label{fig:performance}
%\end{figure*}

\begin{table*}
\small
\begin{center}
 \caption{Speech synthesis quality estimation for four testing conditions on VCTK data. (S: softmax, AS: angular-softmax). Highlighted values correspond to best system based on `Avg' across conditions. }
 \begin{tabular}{|ll|ccccc||ccccc||ccccc|}
\hline
& & \multicolumn{5}{c||}{Estimated MOS}  & \multicolumn{5}{c||}{Speaker Similarity} & \multicolumn{5}{c|}{Intelligibility (WER)}\\
\hline
Method & & C1 & C2 & C3 & C4 & Avg & C1 & C2 & C3 & C4 & Avg & C1 & C2 & C3 & C4 & Avg\\
\hline\hline
Natural Speech      & -- &4.1 &3.9 & 3.8& 3.6 &3.8 & --& --& --&--&--  & 9.0& 10.6& 8.4&8.2&9.0\\\hline
VQ-VAE       & -- & 3.6&3.5 &3.7 &3.3&3.5   & 0.94& 0.94& 0.33&0.46&0.66  &40.4 & 49.1& 85.5&87.9&65.6\\\hline
+ Global VQ  & -- &2.3 &2.4 &2.1 &2.2&2.2   & 0.42& 0.45& 0.54&0.59&0.50  & 83.8& 82.4& 87.7&74.5&82.1\\\hline
\multirow{2}{*}{+ Speaker label} & S &4.1 &3.9 &3.8 &3.8&3.9   &\cellcolor{LightCyan}\textbf{0.89} &\cellcolor{LightCyan}\textbf{0.88} &\cellcolor{LightCyan}\textbf{0.91} &\cellcolor{LightCyan}\textbf{0.91} &\cellcolor{LightCyan}\textbf{0.89} &25.8 & 40.2& 27.7&30.8&31.1\\
& AS & 3.9& 3.8& 3.7&3.7&3.7   &0.87 &0.86 &0.87 &0.87&0.87  & 30.4&42.3 & 30.3&29.2&33.5\\\hline
\multirow{2}{*}{+ Adversarial loss} & S &\cellcolor{LightCyan}\textbf{4.1} &\cellcolor{LightCyan}\textbf{3.9} &\cellcolor{LightCyan}\textbf{3.9}&\cellcolor{LightCyan}\textbf{3.9}&\cellcolor{LightCyan}\textbf{4.0}  & \cellcolor{LightCyan}\textbf{0.89}&\cellcolor{LightCyan}\textbf{0.89} &\cellcolor{LightCyan}\textbf{0.89} &\cellcolor{LightCyan}\textbf{0.90}&\cellcolor{LightCyan}\textbf{0.89}  & \cellcolor{LightCyan}\textbf{26.3}& \cellcolor{LightCyan}\textbf{34.9}& \cellcolor{LightCyan}\textbf{22.5}&\cellcolor{LightCyan}\textbf{26.8}&\cellcolor{LightCyan}\textbf{27.6}\\
& AS &4.1 &4.0 &4.0 &3.8&3.9   &0.84 &0.82 & 0.84& 0.84&0.84 & 32.0&39.2 & 28.3&35.6&33.7\\\hline
%\hline
 \end{tabular}
 \label{tab:quality}
 \end{center}
 \vspace{-4mm}
\end{table*}

\section{Experiments}
We used English audio from the VCTK corpus~\cite{yamagishi2019vctk} and followed the data preparation steps from scripts provided by~\cite{zhao2020improved}, including quantization and normalization. All of our audio was downsampled to 16 kHz. We created a training set of 100 speakers, and a held-out set of 10 speakers. Further, we compiled four testing conditions, each with 10 speakers and 8 utterances per speaker: (\textit{C1}) seen speakers/seen texts, (\textit{C2}) seen speakers/unseen texts, (\textit{C3}) unseen speakers/seen texts, and (\textit{C4}) unseen speakers/unseen texts. All of the VQ-VAE variant systems were trained using similar initialization and configuration from~\cite{zhao2020improved}. However, for our dual-encoder systems we used our fully-trained original \textbf{VQ-VAE} model as a warm-up. This was done to reduce overall training time. The original \textbf{VQ-VAE} was trained to 800k steps, while the dual-encoder systems were trained to an additional 800k steps, and the best model was selected based on lowest validation loss. 

We also fine-tuned each VQ-VAE variant to the TIMIT corpus~\cite{garofolo1993darpa} for our later phone recognition task. The train set had 462 speakers and the test set had 168 held-out speakers (10 utterances each). We partitioned the default train set into half (5 utterances per speaker) and used one portion for fine-tuning our models. The remaining train/test partitions were used for our downstream tasks. To perform fine-tuning, we froze the gradients of all speaker-related components, and fine-tuned only the phone-related components to 400k steps, selecting the best model based on lowest validation loss.

\section{Objective Evaluation}
Estimated mean-opinion scores (MOS) of natural and generated speech\footnote{Audio samples: \url{https://rhoposit.github.io/icassp2021}} were obtained automatically using a pre-trained model\footnote{\url{https://github.com/rhoposit/MOS_Estimation2}} from different TTS and VC systems from~\cite{williams2020comparison} and are shown in Table~\ref{tab:quality}. The slightly lower MOS scores for natural speech highlights a byproduct of using fully-automated methods for speech quality evaluation, and we note that natural speech and some proposed methods appear to be very close in quality. Our semi-supervised systems achieved higher estimated MOS scores compared with the original \textbf{VQ-VAE}. Speaker similarity was determined by calculating the cosine distance between extracted utterance-level \textit{x}-vectors\footnote{\url{https://kaldi-asr.org/models/m7}} for natural and synthetic speech. The \textbf{VQ-VAE} system has high speaker similarity for seen conditions, but falls sharply in unseen conditions. Our semi-supervised and adversarial approaches are more consistent across conditions. Intelligibility was estimated by measuring word error rate (WER) from ASR~\cite{morris2004and}. We used the Watson Speech Recognition API\footnote{\url{https://www.ibm.com/cloud/watson-speech-to-text}}. The original \textbf{VQ-VAE} system achieved a nominal WER score in seen conditions, however the error increased significantly for unseen conditions. The self-supervised system (\textbf{+Global VQ}) was unable to produce intelligible output across all conditions. The semi-supervised system (\textbf{+Adversarial loss}) with softmax maintained the lowest WER across all conditions, including the unseen conditions. Overall, we did not find significant improvements from using angular-softmax. 
%Word Error Rate\footnote{\url{https://pypi.org/project/jiwer/}}
%\footnote{\url{https://github.com/rhoposit/MOS_Estimation2}}

\begin{figure}
  \centering
   \includegraphics[width=8.5cm]{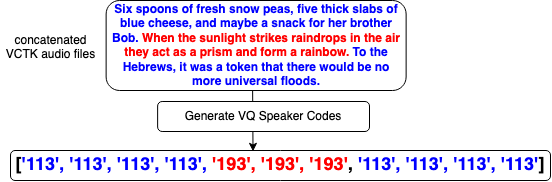}
   \caption{VQ speaker codes are generated at each sliding window, for a given audio file that contains two different speakers. The codes determine which regions of speech belong to speaker A (\textbf{{\color{blue}113}}) versus speaker B (\textbf{{\color{red}193}}).}
   \label{fig:diarize}
\end{figure}

\section{Speaker Diarization Task}
The goal of speaker diarization is to annotate which regions of speech belong to different speakers. We simulated a simple diarization task by concatenating audio files from speakers in our VCTK test conditions, so that there were 2 speakers and 3 turns per audio file, as shown in Figure~\ref{fig:diarize}. We used a 2s sliding window (250ms overlap) to obtain short segments of audio. For our VQ-based methods, a speaker code was obtained for each segment. Single-speaker audio VQ reference codes were used as a guide as well as knowledge that 2 speakers were present in each file. For example, if the speaker code \textit{113} for an audio segment belonged to \textit{speaker A}, then we labeled that region as \textit{speaker A}. In all systems, sometimes two different speakers were mapped into the same VQ code during training. In the case where speaker could not be determined, we chose \textit{speaker A} or \textit{speaker B} randomly. We compared our predictions with the Track1 \textit{x}-vector baseline from the 2019 DIHARD challenge. It used PLDA scoring and agglomerative hierarchical clustering~\cite{sell2018diarization}. Diarization clustering for the baseline method was tuned on the DIHARD dev set, and evaluated on each of our VCTK test conditions using knowledge of 2 speakers per audio file. We report diarization error rate (DER) in Table~\ref{tab:diarization}. The DER score measures the proportion of time in an audio file wherein: non-speech is incorrectly labeled as speech (false alarm), a speech region is incorrectly labeled as non-speech (miss), and an incorrect speaker label is generated (error). The \textbf{+Speaker label} and \textbf{+Adversarial loss} systems performed better than the strong \textit{x}-vector baseline on average. The self-supervised system (\textbf{+Global VQ}) did not learn a diverse VQ space, and we observed the same code was generated for all speakers (sometimes referred to as "codebook collapse"). 

\begin{table}[]
\small
\begin{center}
 \caption{Speaker diarization error (DER) scores on concatenated VCTK audio. (S: softmax, AS: angular-softmax).}
 \begin{tabular}{|ll|ccccc|}
\hline
& & \multicolumn{5}{c|}{Condition}\\
\hline
Method & & C1 & C2 & C3 & C4 & Avg\\
\hline\hline
\textit{x}-vector & & 24.3 &44.6 & 27.4& 46.7& 35.8\\\hline\hline
VQ-VAE &  & -- & -- & -- & -- & -- \\\hline
+ Global VQ & & 44.4& 39.1& 44.7& 39.6& 42.0\\\hline
\multirow{2}{*}{+ Speaker label} 
& S & 32.4 & 32.2& 31.0& 33.1& 32.2\\
& AS & 34.6&35.9 & 36.4& 35.9&35.7\\\hline
\multirow{2}{*}{+ Adversarial loss}
& S & \cellcolor{LightCyan}\textbf{32.2}& \cellcolor{LightCyan}\textbf{32.3}& \cellcolor{LightCyan}\textbf{30.5}&\cellcolor{LightCyan}\textbf{32.9} & \cellcolor{LightCyan}\textbf{31.9}\\
& AS & 37.2& 35.6& 36.1& 35.2& 36.0\\\hline
 \end{tabular}
\label{tab:diarization}
\end{center}
\vspace{-5mm}
\end{table}

\section{Phone Recognition Task}
We conducted a phone recognition task using TIMIT data and a simple LSTM encoder-decoder architecture. The encoder input was either a string of discrete sub-phone VQ codes, or audio features, and the output was a string of the recognized phones. We used the ESPNet\cite{inaguma-etal-2020-espnet} Kaldi recipe for CMU AN4 ASR~\cite{DBLP:journals/corr/KimHW16} and a single-layer LSTM model with 64 units, trained to 100 epochs, selected the best model based on CTC loss without attention ($mtlalpha = 1.0$), and decoded with beam size 20. In TIMIT there were 63 unique phone types. For our VQ methods, first we applied our fine-tuned models to extract VQ code sequences for each audio file. The lengths of sub-phone VQ code sequences depend on the waveform downsampling factor (DSF) from the VQ-VAE encoder (we used DSF$=64$). Likewise, the sub-phone codewords varied by system since each version learned a slightly different VQ sub-phone space. Table~\ref{tab:asr} shows the number of unique codewords utilized in each system, as well as the mean phone error rate (PER) averaged across all speakers for phone substitution, insertion, deletion, and overall (Sub, Ins, Del, Total). The VQ-VAE systems offer error reductions for insertion and deletion, compared to the strong audio baseline which has fewer substitutions overall. The \textbf{+ Speaker label AS} system had better overall performance over the \textbf{+Global VQ} system. These results show it is possible to add global-level speaker components to VQ-VAE without sacrificing quality of sub-phone representations.

\begin{table}
\small
\begin{center}
 \caption{Phone error rate (\% PER) on TIMIT from sub-phone VQ codes or audio. (S: softmax, AS: angular-softmax).}
 \begin{tabular}{|ll|c|c|c|c|c|}
\hline
& & \# VQ& \multicolumn{4}{c|}{\% PER}\\
Method & & Codes & Sub& Ins& Del & Total\\
\hline\hline
Audio Baseline & & -- & 13.8&9.4&7.4&30.6 \\ \hline\hline
VQ-VAE & & 140 & \cellcolor{LightCyan}\textbf{26.6}&\cellcolor{LightCyan}\textbf{8.3}& \cellcolor{LightCyan}\textbf{6.0}&\cellcolor{LightCyan}\textbf{40.9}\\\hline
+ Global VQ & & 119 & 28.0 &7.7&6.4& 42.1 \\ \hline
\multirow{2}{*}{+ Speaker label}
& S & 139&28.1 &9.6&5.8& 43.4\\
& AS & 138& \cellcolor{LightCyan}\textbf{27.6}&\cellcolor{LightCyan}\textbf{8.0}& \cellcolor{LightCyan}\textbf{6.3}& \cellcolor{LightCyan}\textbf{41.9}\\\hline
\multirow{2}{*}{+ Adversarial loss} 
& S & 176&28.0 &8.6&6.5& 43.1\\
& AS & 154& 30.4&9.6&6.5& 46.5\\\hline
 \end{tabular}
\label{tab:asr}
\end{center}
\vspace{-5mm}
\end{table}

\section{Discussion}
We have presented an approach toward learning disentangled representations of global speaker characteristics and sub-phone content that are also useful for downstream tasks\footnote{Code/models: \url{https://github.com/rhoposit/icassp2021}}. Adding a semi-supervised VQ component with either speaker labels or adversarial loss improves objective estimates of speech synthesis quality, intelligibility, and speaker similarity from the original VQ-VAE in unseen conditions. Additional work on automated MOS scoring research should consider incorporating natural speech as a reference. Our simulated speaker diarization task shows that speaker VQ codes could be adapted for speaker diarization. From our phone recognition experiments, we found that none of the systems utilized the full VQ space (512 codebooks) so future experiments should determine the optimal codebook sizes, which may depend on the language and diversity of speakers. We plan to explore additional disentanglement by including an F0 modeling component. Speaker VQ codes can also be used conveniently for voice conversion, as this would not require external representations (such as \textit{x}-vector or LDE). Synthesizing speech from discrete VQ codes may facilitate the ability to control expressiveness in text-to-speech synthesis. For example, a particular sequence of VQ codes could be used to control prosody alongside other signal factors.

% References should be produced using the bibtex program from suitable
% BiBTeX files (here: strings, refs, manuals). The IEEEbib.bst bibliography
% style file from IEEE produces unsorted bibliography list.
% -------------------------------------------------------------------------
\bibliographystyle{IEEE}
\bibliography{refs}

\end{document}